\documentclass[aps,prl,twocolumn,showpacs]{revtex4}

\usepackage{graphicx}

\begin{document}

%\draft

\title{Proximity-induced sub-gaps in Andreev billiards}

\author{ J. Cserti\thanks{e-mail: cserti@galahad.elte.hu},
A. Korm\'anyos, Z. Kaufmann, J. Koltai}
\affiliation{Department of Physics of Complex Systems, 
E{\"o}tv{\"o}s University
\\ H-1117 Budapest, P\'azm\'any P{\'e}ter s{\'e}t\'any 1/A, Hungary}
\author{C. J. Lambert\thanks{e-mail: c.lambert@lancaster.ac.uk}}
%\author{C. J. Lambert\thanks{e-mail: c.lambert@lancaster.ac.uk}}
\affiliation{Department of Physics, Lancaster University,
Lancaster, LA1 4YB, UK}
%\date{\today}

%\wideabs{

\begin{abstract}

We examine the density of states of an Andreev billiard and show
that any billiard with a finite upper cut-off in the path length
distribution $P(s)$ will possess an energy gap on the scale of the
Thouless energy. An exact quantum mechanical calculation for
different Andreev billiards gives good agreement with the
semiclassical predictions when the energy dependent phase shift
for Andreev reflections is properly taken into account.  Based on
this new semiclassical Bohr-Sommerfeld approximation of the
density of states, we derive a simple formula for the energy gap.
 We show that the energy gap,
in units of Thouless energy, may exceed the value predicted
earlier from random matrix theory for chaotic billiards.

\end{abstract}


\maketitle

%}

Studies of sub-gap transport in hybrid superconductors are an
important starting point for the design and simulation of
superconducting nanoscale devices. For a given inhomogeneous
structure, a fundamental question is the existence or otherwise of
a finite density of states at the Fermi energy. This is important
experimentally, since the sub-gap spectrum determines the
tunneling conductance of an N-S contact. The ability to address
this question is also important theoretically, since a well-posed
problem of this kind provides a testing ground for complementary
(and occasionally competing) theoretical techniques.

\begin{figure}[hbt]
\includegraphics[scale=0.5]{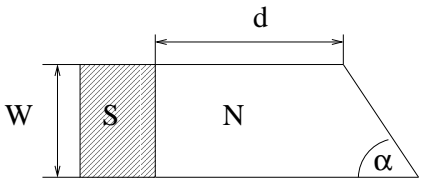}
%\centerline{\leavevmode \epsfxsize=7cm \epsffile{wedge-geo.eps }}
\caption{Geometry. \label{geo-fig}}
\end{figure}

One important class of structures, for which a general analysis
might be forthcoming, are known as Andreev billiards. These are
formed when  a classically-chaotic normal dot is placed in contact
with a superconductor
\cite{Kosztin,Melsen,Altland,Lesovik,Nazarov,Heny,Richter1,Richter2}.
Consider a ballistic two dimensional normal dot of area $A$, with
the mean level spacing of the isolated normal system $\delta
=\frac{2\pi \hbar^2}{mA}$ at the Fermi energy $E_F$. If a
superconductor of width $W$ and bulk order parameter $\Delta$ is
placed in contact with such a billiard (see Fig.~\ref{geo-fig}),
then the question of interest is whether or not an energy gap on
the scale of the Thouless energy
$E_T=\frac{M\delta}{4\pi}$~\cite{note_1}, exists in the sub-gap
spectrum $E <\Delta$, when $\delta << E_T < \Delta$. The number
of open channels in the S region is the integer part of
$M=\frac{k_{\rm F}W}{\pi}$, and the energy levels of the Andreev
billiards  are the positive eigenvalues $E$ (measured from the
Fermi energy) of the Bogoliubov-de Gennes equation\cite{BdG-eq}.

An initial study of this problem based on random matrix
theory\cite{Melsen}, concluded that a classically chaotic
billiard possesses an energy gap, whereas an integrable system is gapless.
As a consequence it was suggested that the existence of such a gap could be
used to distinguish between integrable and chaotic systems.
Later it was shown that billiards with classically mixed phase space  
do possess a smaller gap 
comparing to the fully chaotic systems\cite{Heny}.
Studying the existence of the gap for chaotic billiards 
Ihra et al.\ \cite{Richter1} have drawn the attention to the role of the
non-diagonal terms of the scattering matrix which could explain the
disagreement between the random matrix theory and the Bohr-Sommerfeld
semiclassical approximation. The non-universal feature of the
excitation spectra of the Andreev billiards has been recently studied
by Ihra and Richter in Ref.\ \cite{Richter2}.

\begin{figure}[hbt]
\includegraphics[scale=0.5]{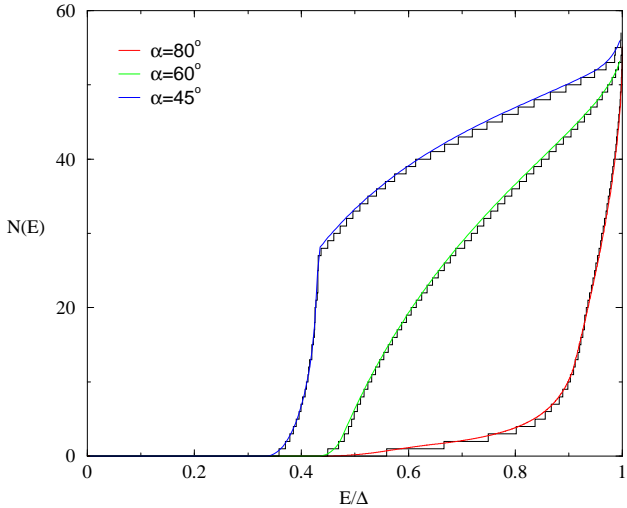}
%\centerline{\leavevmode \epsfxsize=7cm \epsffile{fig2.eps }}
\caption{The integrated density of states, $N(E)$ as a function of $E$ (in
units of $\Delta$) for $d=0$ and angle
$\alpha = 80^{\circ}, 60^{\circ}, 45^{\circ}$.
For each angle $\alpha$ the stair type lines correspond
to the exact diagonalization of the Bogoliubov-de Gennes equation,
while the solid lines are obtained by using Eq.~(\ref{N-eq}).
The parameters are $M=55.5$, $\Delta/E_F = 0.015$.
\label{N-zero_d-fig}}
\end{figure}

\begin{figure}[hbt]
\includegraphics[scale=0.45]{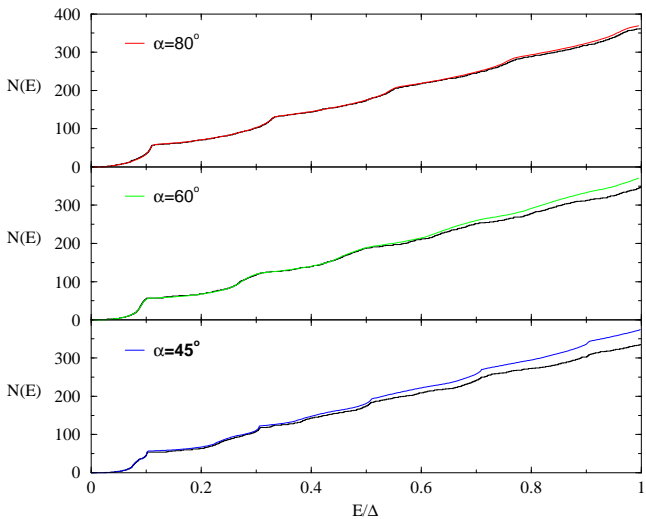}
%\centerline{\leavevmode \epsfxsize=7cm \epsffile{fig3.eps }}
\caption{The integrated density of states, $N(E)$ as a function
of $E$ (in units of $\Delta$) for $d \ne 0$ and angle $\alpha =
80^{\circ}, 60^{\circ}, 45^{\circ}$. In each case, $d$ varies
such that the area of the normal dot is fixed at $A=5W^2$ for the
angles $\alpha=80^{\circ}, 60^{\circ}, 45^{\circ}$. This choice
results in approximately the same number of states below $\Delta$
for different values of $\alpha$. The meaning of the lines and the
parameters are the same as those in Fig.~\ref{N-zero_d-fig}. 
\label{N-nonzero_d-fig}}
\end{figure}

The aim of this Letter is to identify for the first time, a
pseudo-integrable  billiard which does possess an energy gap. 
Moreover,  this gap can be larger than that predicted for
chaotic billiards on the scale of $E_T$. To this end,
we study the quasi-particle spectrum of the ballistic structure shown
in Fig.~\ref{geo-fig}, for different values of the angle $\alpha$
and length $d$. Our key result is illustrated
in Fig.~\ref{N-zero_d-fig}, which shows that for $d=0$,
the counting function or integrated density of states
$N(E)=\int_0^E\, n(E^\prime)\, dE^\prime$ (where $n(E)$ is the density of
states for the Andreev billiard) exhibits an
energy gap $E_{\rm gap}$ varying between
$E_{\rm gap} \approx 0.5 E_T$ and $E_{\rm gap}\approx 1.5 E_T$
as $\alpha$ varies from  $80^{\circ}$ to $45^{\circ}$
(for the details of the calculation see below).
In contrast, as shown in Fig.~\ref{N-nonzero_d-fig}, for $d \ne
0$, no such energy gap exists.
To understand this result, we start from the formula relating
the integrated density of states $N(E)$ to the path length distribution
$P(s)$ (which is the classical probability
that an electron entering the billiard at the N-S contact exits
after a path length $s$) derived by using the Bohr-Sommerfeld approximation,
\begin{equation}
N\left(E \right) = M \sum_{n=0}^\infty \,
\int_{s_n\left(E \right)}^\infty \, P(s)\, ds,
\label{N-eq}
\end{equation}
where
\begin{equation}
s_n\left(E \right) = \frac{\left(n+
\frac{1}{\pi}\arccos\frac{E}{\Delta}\right)\pi}{E/\Delta}\,\, \xi_0.
\label{s_eps-eq}
\end{equation}
Here $\xi_0=\frac{\hbar v_{\rm F}}{\Delta}
= W\, \frac{2}{\pi M}\, \frac{E_F}{\Delta}$
is the coherence length and
the distribution $P(s)$ is normalized to one,
$\int_0^\infty\, P(s)\, ds =1$.
The density of states $n(E)=\frac{dN(E)}{dE}$ can be found
from Eq.~(\ref{N-eq}) yielding
\begin{equation}
n\left(E \right) =
\frac{M}{E} \sum_{n=0}^\infty \,
P\left(s_n\left(E \right) \right)
\left[\frac{\xi_0}{\sqrt{1-{\left(E/\Delta\right)}^2}} +
s_n\left(E \right)\right].
\label{n-eq}
\end{equation}
This expression reduces to that of derived in a different way
by Melsen et al.\ \cite{Melsen,Heny}, Lodder and Nazarov\cite{Nazarov},
Ihra et al.\ \cite{Richter1}, in the limit
$E << \Delta$ and $\xi_0^2 << A$,
but more generally incorporates an energy dependent path length
correction~\cite{note_2}.

To test the semiclassical expression, we  perform an
exact (numerical) diagonalization of the Bogoliubov-de Gennes
equation by  matching the wave functions at the N-S interface.
This results in a secular equation including the scattering matrix,
$S_0(E)$ of the normal billiard opened at the N-S interface~\cite{elsewhere}:
\begin{equation}
\det \left[1-e^{-2i\arccos\frac{E}{\Delta}}\,
S_{\rm eff}(E)\, S_{\rm eff}^*(-E) \right] = 0,
\label{secular-eq}
\end{equation}
where
\begin{eqnarray}
S_{\rm eff}(E) &=&
{\left[Q(E)+K(E)D(E)\right]}^{-1}  \times
\nonumber \\
&& \left[Q^*(E)-K(E)D(E)\right],
\label{Seff-eq}\\[2ex]
D(E)&=&\left[1-S_0(E)\right]{\left[1+S_0(E)\right]}^{-1}. \nonumber
\end{eqnarray}
Here $Q$ and $K$ are diagonal matrices with elements
$Q_{nm}(E)=i\, \delta_{nm}\,q_n(E)$ and $K_{nm}(E)=\delta_{nm}\,
k_n(E)$, where $q_n (E)=k_F \sqrt{1 + i\,
\frac{\sqrt{\Delta^2-E^2}}{E_F}-\frac{n^2}{M^2}}$ are the
transverse wavenumbers of the electron in the S region and
$k_n(E)=k_F\sqrt{1+\frac{E}{E_F}-\frac{n^2}{M^2}}$ are the
transverse wavenumbers of the electron in the S region when
$\Delta=0$.  It is assumed that the Fermi wavenumber, $k_F =
\sqrt{2mE_F/\hbar^2}$ is the same in the S and N regions. All
the matrices are $M$ by $M$ dimensional. Finally, $S_0$ was
calculated  using Bessel functions in the wedge part of the
normal region, and  including evanescent modes. In a different
context, the same type of normal billiard was studied by Kaplan
and Heller~\cite{Heller}. A secular equation similar to
Eq.~(\ref{secular-eq}) was derived by Beenakker~\cite{Carlo_2_3}
for SNS systems but there instead of $S_{\rm eff}(E)$ the
scattering matrix $S_0(E)$ of the normal billiard appears. In
this sense, the matrix $S_{\rm eff}(E)$ can be regarded as an
effective scattering matrix. Note that in Andreev 
approximation\cite{Colin-review} $q_n\approx k_n$, and one then
finds $S_{\rm eff}(E)=S_0(E)$. In contrast the secular
equation~(\ref{secular-eq}) for Andreev billiards is valid outside
Andreev approximation. Using the unitarity of $S_0(E)$, one can
show that the following equation
\begin{eqnarray}
\det \left(\rm Im \left\{
e^{-i\arccos\frac{E}{\Delta}} \,
\left[Q(E)+K(E)D(E)\right] \times  \right. \right.
\nonumber && \\
\left. \left.
{\left[Q(E)+K(E)D^*(-E)\right]}^{-1}
\right\}
\right) &=& 0,
\label{secular-real-eq}
\end{eqnarray}
has the same zeros as those of Eq.~(\ref{secular-eq}), and is a
more suitable form for numerical calculations (the left hand side
of Eq.~(\ref{secular-real-eq}) is a real function instead of a
complex one).

Figures~\ref{N-zero_d-fig} and \ref{N-nonzero_d-fig} show both
exact results and an evaluation of the semiclassical expression
of $N(E)$ given by Eq.~(\ref{N-eq}) for Andreev billiards shown
in Fig.~\ref{geo-fig} with different parameters~\cite{note_3}.
One can see that the results of the two methods are in good
agreement over the entire energy range $E<\Delta$. It is important
to note that inclusion of the energy dependence of the phase shift
was essential to obtain such a good agreement. A small deviation
can be seen only at energies close to the value of $\Delta$ and
for $d\ne 0$.

Both the semiclassical and the exact calculations reveal the
presence of an energy gap when $d=0$, while no gap appears in
case of $d\ne 0$. The origin of this result can be traced to the
existence of a finite upper cut-off $s_{\rm max}$ in the path
length distribution $P(s)$. From Eq.~(\ref{s_eps-eq}) it is
obvious that if the path length of the electron has a maximum
value, then the energy spectrum will possess a gap. The condition
for determining $E_{\rm gap}$ is $s_n(E) > s_{\rm max}$ for $n\ge
0$. Expanding the arccos term in Eq.~(\ref{s_eps-eq}) to first
order in $E/\Delta$, one finds that the energy gap is given by the
following simple equation:
\begin{equation}
\frac{E_{\rm gap}}{E_{\rm T}} = \pi^2 \,
\frac{A}{Ws_{\rm max}}\, \frac{1}{1+\xi_0/s_{\rm max}},
\label{gap-eq}
\end{equation}
where $A$ is the area of the billiard and $W$ is the width of the
superconductor lead.

\begin{figure}[hbt]
\includegraphics[scale=0.45]{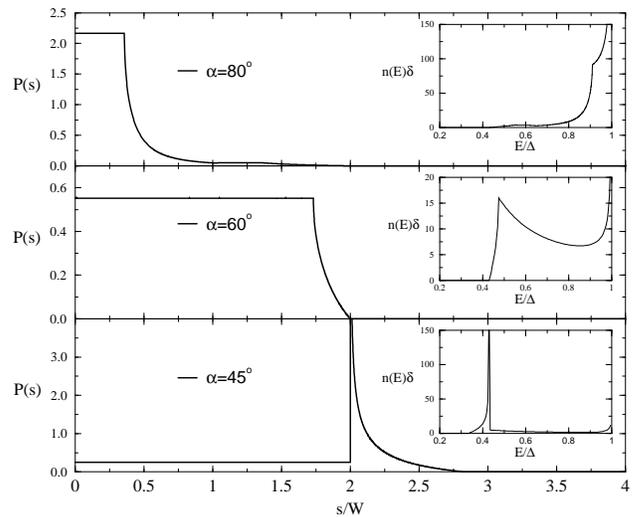}
%\centerline{\leavevmode \epsfxsize=7cm \epsffile{fig4ff.eps }}
\caption{The path length distribution $P(s)$ for Andreev billiard of
Fig.~\ref{geo-fig} when $d=0$
for $\alpha=80^{\circ}, 60^{\circ}, 45^{\circ}$.
Insets show the corresponding semiclassical density of states $n(E)$
(in units of $1/\delta$) using Eq.~(\ref{n-eq}).
The same parameters was used as those in Fig.~\ref{N-zero_d-fig}.
\label{Ps-fig}}
\end{figure}

As examples, Fig.~\ref{Ps-fig} shows $P(s)$ for
$\alpha=80^{\circ}, 60^{\circ}, 45^{\circ}$ along with the
corresponding semiclassical density of states $n(E)$ given by
Eq.~(\ref{n-eq}). From the figure it is seen that the upper
cut-off $s_{\rm max}$ of $P(s)$ equals to $2W$ for $\alpha
=80^{\circ}, 60^{\circ}$, and $s_{\rm max} = 2\sqrt{2}\,W$ for
$\alpha = 45^{\circ}$. It is interesting to note that for $\alpha
=45^\circ$ (more generally for $\alpha = 90^{\circ}/k$ with an
integer $k>0$) the density of states $n(E)$ has a 
singularity at some
energy $E<\Delta$ while for other values of $\alpha$ no
pronounced singularities exist. In Bohr-Sommerfeld approximation,
it can be shown that this singularity is related to the
singularity of the path length distribution $P(s)$. The classical
trajectories of the electron resulting in a singularity 
for $\alpha =45^\circ$ are 
similar to those moving in a lead with length $d+W$ and 
$\alpha =90^\circ$. The nature of the singularities of the
density of states in this rectangular shape of Andreev billiard 
was studied in Ref.~\onlinecite{box-disk-paper}. 
The reason for the disappearance 
of these singularities for $\alpha \ne 90^\circ/k$ ($k=1,2,3,\ldots$) is
consistent with the points made in this reference.

\begin{figure}[hbt]
\includegraphics[scale=0.45]{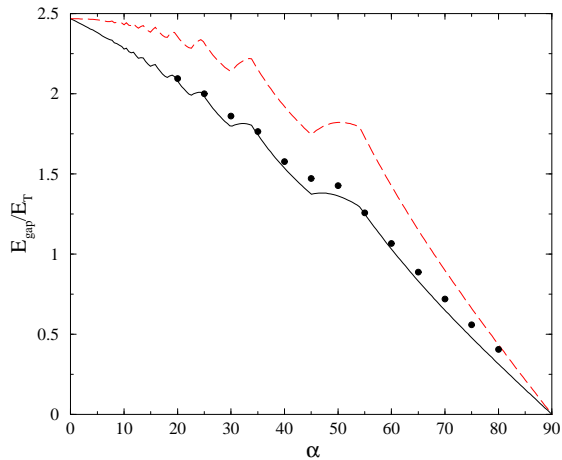}
%\centerline{\leavevmode \epsfxsize=7cm \epsffile{gap-BS.eps }}
\caption{The gap $E_{\rm gap}$ (in units of Thouless energy)
as a function of angle $\alpha$ for Andreev billiards with $d=0$.
The solid and the dashed curves are the semiclassical results
from Eq.~(\ref{gap-eq}) when $M=55.5$, and from Eq.~(\ref{gap-semi-eq})
i.e.\ when $M\rightarrow \infty$, respectively.
The circles are the results of the exact diagonalization when $M=55.5$.
\label{gap-fig}}
\end{figure}

To check the expression (\ref{gap-eq}) we have determined the
energy gap from exact calculations for several angles $\alpha$.
 For the billiard of
Fig.~\ref{geo-fig} with $d=0$, one can show analytically that the
upper cut-off $s_{\rm max}$ is
\begin{equation}
\frac{s_{\rm max}}{W}= \left\{
\begin{array}{ll}
-2\, \frac{\sin k\alpha \sin \left[\left(k+1\right)\alpha\right]}
{\sin\alpha \cos \left[\left(2k+1\right)\alpha\right]}
%\frac{1-\cos \alpha/\cos \left[\left(2k+1\right)\alpha\right]}{\sin\alpha}
& \mbox{if \,\,
$\frac{\pi/2}{k+1} < \alpha \le \frac{\pi/2}{k+\frac{2}{3}}$}, \\[1.2ex]
%\frac{\sqrt{2\left(1-\cos 2k\alpha \right)}}{\sin \alpha}
2\, \frac{\sin k\alpha}{\sin \alpha}
& \mbox{if \,\,
$\frac{\pi/2}{k+\frac{2}{3}} < \alpha \le \frac{\pi/2}{k}$},
\end{array}
\right.
\label{smax-exp}
\end{equation}
where $k=1,2,3,\ldots .$
Using Eqs.~(\ref{gap-eq}) and (\ref{smax-exp})
the gap is plotted as a function of the angle $\alpha$
in Fig.~\ref{gap-fig} together with the exact results.
The numerical results (full circles in Fig.~\ref{gap-fig}) agree
very well with the predictions of Eq.~(\ref{gap-eq}). One can see
that decreasing $\alpha$ the energy gap in units of Thouless
energy increases and tends to a finite value as $\alpha
\rightarrow 0$. From Eq.~(\ref{smax-exp}) and Eq.~(\ref{s_eps-eq})
it can be seen that $s_{\rm max}$ increases with decreasing
$\alpha$, whereas $E_{\rm gap}/\Delta$ decreases. The Thouless
energy $E_T$ also decreases, since the mean level spacing $\delta$
becomes smaller for larger area $A$ of the normal region. The two
effect together result in a finite value of $E_{\rm gap}/E_T$ at
$\alpha = 0$.

In the limit $\xi_0 << s_{\rm max}$ (for example, in the semiclassical
limit of large $M$), the expression~(\ref{gap-eq}) of the gap
reduces to
\begin{equation}
\frac{E_{\rm gap}}{E_{\rm T}} = \pi^2 \,
\frac{A}{Ws_{\rm max}}.
\label{gap-semi-eq}
\end{equation}
In Fig.~\ref{gap-fig} the dashed curve is the result from this
formula. One can see that for all angles $\alpha$ this limiting
result for the energy gap is larger than the finite $M$ value.
Note that the area  is $A=\frac{1}{2} \,W^2\,\cot \alpha $, and
therefore, $E_{\rm gap}/E_{\rm T}$ is {\em only} a function of
$\alpha$ and independent of $W$ when $\xi_0 << s_{\rm max}$. It
can also be shown that $E_{\rm gap}/E_{\rm T} \rightarrow \pi^2/4
\approx 2.47$ as $\alpha \rightarrow 0.$ Finally, it is worth
mentioning that for Andreev billiard of Fig.~\ref{geo-fig} when
$d=0$ the energy gap
can be much larger for small angle $\alpha$ than the value of
$0.6 E_T$ predicted by Melsen et al.~\cite{Melsen} for the chaotic
billiard.

\begin{figure}[hbt]
\includegraphics[scale=0.45]{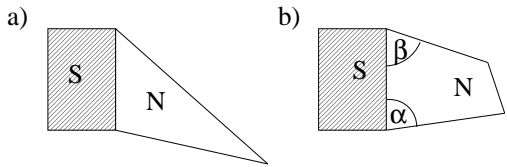}
%\centerline{\leavevmode \epsfxsize=85mm \epsffile{other-dot.eps }}
\caption{A triangle (a), and quadrangle (b) shape of normal billiard 
having a finite upper cut-off of the length distribution $P(s)$.
\label{other_dots-fig}}
\end{figure}

It is interesting to mention that besides the billiard studied 
in this work (see Fig.~\ref{geo-fig} for $d=0$) 
there are a number of different shapes of 
normal dots in which $P(s)$ exhibits finite upper cut-off resulting in a
sub-gap. Examples for such dots are shown in
Fig.~\ref{other_dots-fig} (provided $\alpha +\beta < 180^{\circ}$ 
in the case of quadrangle).

In conclusion, we have shown that a billiard with a finite upper
cut-off in the path length distribution $P(s)$ will possess an
energy gap in the density of states on the scale of the Thouless
energy. By including the energy dependent phase shift of the
Andreev reflection, a new expression for the density of states
has been given within the frame work of the semiclassical
Bohr-Sommerfeld approximation. We have also derived a formula for
the energy gap. To check these results we have performed an exact
diagonalization of the Bogoliubov-de Gennes equation for
different Andreev billiards. The results of the two methods
agrees very well both for the integrated
density of states (even for energy levels close to the value of
$\Delta$) and for the energy gap. Finally we have shown that the
energy gap on the scale of the Thouless energy can be much larger
than the value $0.6 E_T$ predicted from random matrix theory\cite{Melsen} 
for chaotic billiards.

One of us (J. Cs.) gratefully acknowledges very helpful discussions
with C. Beenakker. 
This work was supported by the EU.\ RTN within the programme
``Nanoscale Dynamics, Coherence and Computation'',
the Hungarian  Science Foundation OTKA  TO25866 and TO34832.
One of us (Z. K.) would like to thank the Hungarian
Academy of Sciences for support as a J\'anos Bolyai fellowship.

%\begin{references}


\begin{thebibliography}{10}
\bibitem{note_1}

The same definition for the  Thouless energy is used
as that for example, in Refs.~\onlinecite{Melsen,Heny}.
\bibitem{note_2}
The  $\arccos$ term in Eq.~(\ref{s_eps-eq}) due to
the phase shift $-\arccos(E/\Delta)$
for Andreev reflection at the N-S interface. This energy dependent
correction was approximated by $\pi/2$
in Refs.~\cite{Melsen,Nazarov,Heny,Richter1}
which is strictly valid only at $E=0$.
In this case the first term in the square bracket of Eq.~(\ref{n-eq})
is missing.
\bibitem{elsewhere}
Details of the analytical calculations will be given in a later publication.
\bibitem{note_3}
$P(s)$ was calculated analytically for $d=0$ and
numerically using Monte Carlo methods for $d\ne0$.
\end{thebibliography}
\end{document}